\documentclass{article}[11pt]
\usepackage{algorithmic}
\usepackage{algorithm}
\usepackage{graphicx}
\title{Accurate Demarcation of Protein Domain Linkers based on Structural Analysis of Linker Probable Region}
\author{Vivekanand V. Samant\footnote{Persistent Systems Ltd. Pune} \and Arvind Hulgeri\footnote{Persistent Systems Ltd. Pune} 
\and Alfonso Valencia\footnote{Spanish National Cancer Research Center} \and Ashish V. Tendulkar\footnote{Tata Institute of Fundamental Research}\footnote{Corresponding Author: ashishvt@tifr.res.in}}
\date{}
\begin{document}
\maketitle
\begin{abstract}
In multi-domain proteins, the domains are connected by a flexible unstructured 
region called as protein domain linker. The accurate demarcation of these 
linkers holds a key to understanding of their biochemical and evolutionary 
attributes. This knowledge helps in designing a suitable linker for 
engineering stable multi-domain chimeric proteins. Here we propose a novel 
method for the demarcation of the linker based on a three-dimensional protein 
structure and a domain definition. The proposed method is based on biological 
knowledge about structural flexibility of the linkers. We performed structural 
analysis on a linker probable region (LPR) around domain boundary points of 
known SCOP domains. The LPR was described using a set of overlapping peptide 
fragments of fixed size. Each peptide fragment was then described by geometric 
invariants (GIs) and subjected to clustering process where the fragments 
corresponding to actual linker come up as outliers. We then discover the 
actual linkers by finding the longest continuous stretch of outlier fragments 
from LPRs. This method was evaluated on a benchmark dataset of 51 continuous
multi-domain proteins, where it achieves F1 score of 0.745 ($0.83$ precision 
and $0.66$ recall). When the method was applied on 725 continuous multi-domain proteins, it was able 
to identify novel linkers that were not reported previously. This method can 
be used in combination with supervised / sequence based linker prediction 
methods for accurate linker demarcation.
\end{abstract}
\section{Introduction}
Complex proteins are made up of several domains that work independently or in 
tandem with the neighboring domains to perform the intended functions in the 
cellular processes \cite{gokhale:2000}. The domains are linked by 
means of flexible structures known as domain linkers. The linkers perform a 
key role in cooperative inter-domain interactions, function regulation, 
protein stability, folding rates, and domain-domain orientation 
\cite{gokhale:2000, george:2002}. The linkers are known to 
possess special biochemical properties such as high solvent accessibility and 
a typical amino acid composition, due to their role and location in the 
protein structure. To further our understanding on these fronts, a systematic 
analysis of the known linkers needs to be performed. The progress is hampered 
by lack of availability of known and reliable linkers. For instance, there 
is no database of experimentally characterized linkers something that would 
be of immense importance in such studies. The improved understanding of linkers 
and their biochemical properties is crucial in designing linkers for 
engineering stable multi-domain proteins.

The most reliable and accurate linker demarcation can be obtained using 
protein structure analysis. Crystallographers usually perform such analysis to 
identify domains and linkers while determining the structure of multi-domain 
proteins. However, in many cases, the domain linkers are not reported 
explicitly and we need to employ computational methods to demarcate the 
linkers based on sequence or structure of the protein. State of the art 
sequence based methods \cite{suyama:2003, bae:2005, dumontier:2005}
can be used to identify a list of putative linkers and these need to be processed 
further using available structural features to determine the actual linkers. These 
methods take amino acid sequence as an input and predict domain boundaries and 
linkers using a domain linker index computed from amino acid propensities in the 
known linker region. Miyazaki and co- workers have proposed neural 
network \cite{miyazaki:2002} and support vector machine \cite{ebina:2009} based 
techniques using amino acid propensities to distinguish intra-domain loops from the 
inter-domain ones. Tanaka and co-workers used predicted secondary structure in 
addition to amino acid propensities to identify loops, which are further 
distinguished between linker and non-linker loops. Domain prediction methods 
are also used to predict linkers by carving out a stretch of residues in the 
inter domain region around domain boundary points \cite{galzitskaya:2003, liu:2004}. 
These methods tend to provide multiple linker predictions with liberal allowance 
for the linker boundaries and hence are not very useful for accurate protein 
linker demarcation. Besides, most of these methods are unable to predict helical 
linkers due to their assumption about linkers being loops.

George and Heringa conducted a systematic study of biochemical properties of 
the linkers extracted from three dimensional structures of multi-domain 
proteins \cite{george:2002, george:2002a}. They first identify structural 
domains using Taylor’s method \cite{taylor:1999} and then extract linkers by 
branching out from domain boundaries until the branches become buried within 
the core of the domain or till the branch becomes 40 residues long. This 
method takes into account biochemical properties of linkers for their 
demarcation without using any of the structural features.	It is well 
known that the linkers assume unique structures due to their placement in the 
protein structure \cite{argos:1990, robinson:1998} and this forms the 
basis of our method. The proposed method performs accurate demarcation of 
linkers given a three dimensional structure and its domain definition. It 
first extracts a linker probable region (LPR) around domain boundary point 
and then performs structural analysis of the LPR to demarcate actual linker. 
We perform a rigorous assessment of the method using a benchmark dataset of 
known linkers extracted from the literature.

The rest of the paper is organized as follows: (i) The Method section explains 
the proposed technique in detail, (ii) The Results section presents findings 
and representative linkers demarcated by the proposed method. It also 
reports accuracy of the method and its performance vis-à-vis other state of 
the art methods, (iii) The Discussion section documents key contributions of 
the method.

\section{Proposed method}
The method takes a set of protein structures and the corresponding domain 
definitions as input and identifies the corresponding domain linkers. 
\begin{figure}[h]
\centering
\includegraphics[scale=1]{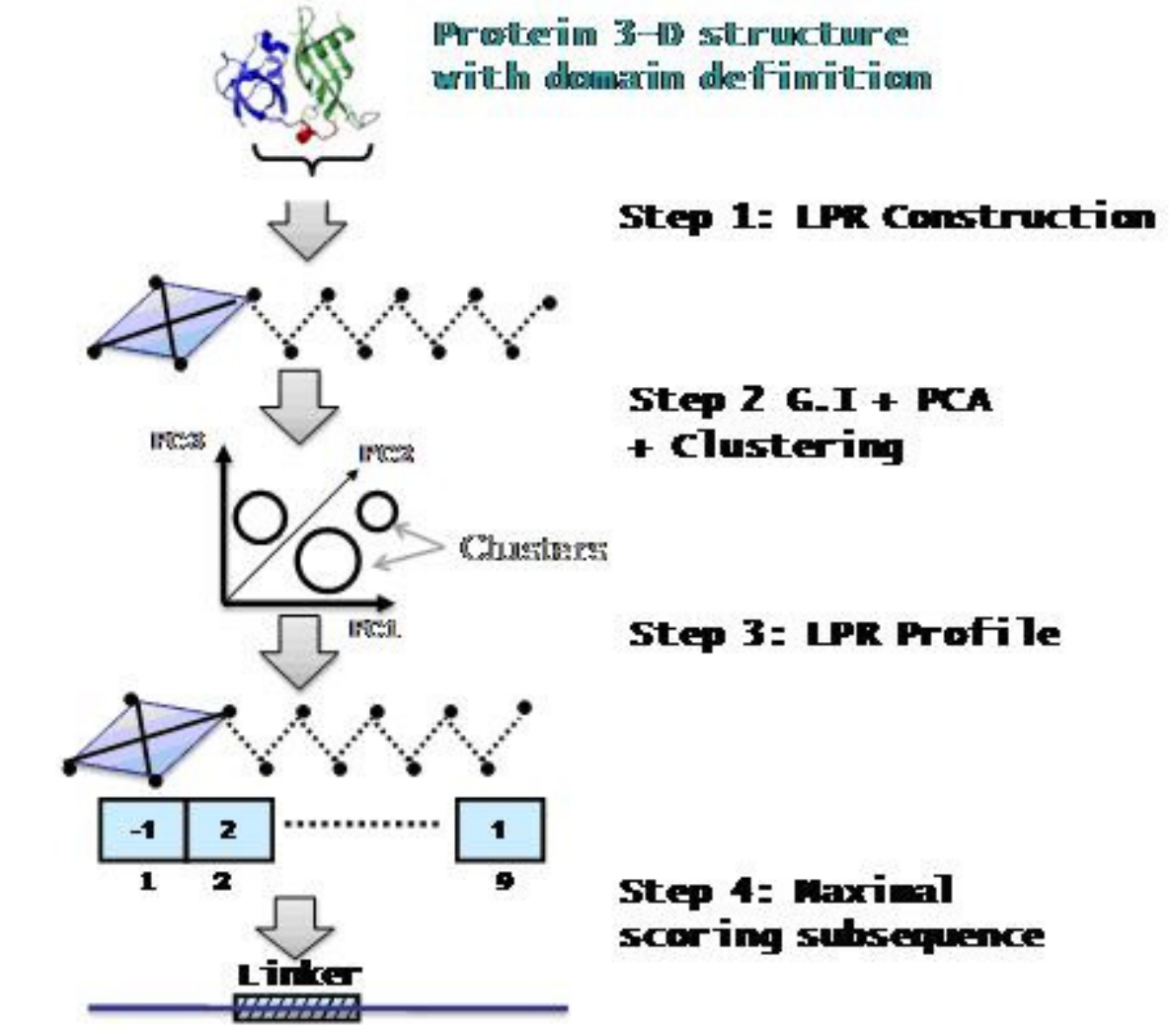}
\caption{\label{fig:Flow-chart}\footnotesize Schematic for protein linker demarcation: 
(i) Based on SCOP domain definition; we identify domain boundary point in 
the protein structure. A twelve residue long linker probable region (LPR) 
is carved out by taking six amino acids each from either domain. 
(ii) LPR is represented with nine overlapping tetrapeptides. (iii) Each 
tetrapeptide is described with a set of 15 geometric invariants (GIs). The 
geometric invariants are standardized to zero mean and unit standard deviation. 
The number of dimensions is reduced via principal component analysis (PCA). 
We select first 8 PCs to represent tetrapeptides. The tetrapeptides are then 
transformed into PC space. (iv) The hierarchical agglomerative clustering is 
performed to identify clusters of similar tetrapeptides. The clusters are 
assigned E-values based on the cluster size distribution. The E-values are 
standardized to zero mean and unit standard deviation, yielding structural 
uniqueness scores (SUS). (v) Based on the membership of a tetrapeptide to a 
particular cluster, we construct LPR structural profiles using SUS of the 
respective clusters. (vi) We perform maximally scoring subsequence 
discovery on top of LPR structural profile to identify the continuous 
stretch with maximum cumulative SUS. This stretch of tetrapeptides 
corresponds to the actual linker.}
\end{figure}

It can be broadly divided into the following four steps, as depicted in Figure 1, 
(i) Construction of linker probable regions (LPRs), (ii) Parameterization of LPRs, 
(iii) Generation of structure profiles by clustering LPRs, and (iv) Demarcation 
of actual linkers by applying dynamic programming on the structure profiles. 
Note that the current version of the method works only with continuous 
multi-domain proteins. The algorithm is given below:

\begin{algorithm}
\algsetup{indent=1em}
\begin{algorithmic} [1]
\REQUIRE 
$\mathcal{S}=\{(S_1, D_1), (S_2, D_2), \ldots, (S_n, D_n)\}$: 
Set of $n$ protein structures with domain definition; and
$k$: Number of positions from one domain to be included in linker probable region
\ENSURE $\mathcal{L} = \{(S_1, L_1), (S_2, L_2), \ldots, (S_n, L_n)\}$.
\FOR{each $(S_i, D_i) \in \mathcal{S}$}
\STATE $R$ = ExtractLPR$(S_i, D_i)$;
\STATE $T$ = DiscretizeLPR($R$);
\STATE $\mathcal{T} = \mathcal{T} + T$
\STATE $\mathcal{R} = \mathcal{R} + R$
\ENDFOR
\STATE $\mathcal{I} = \mathrm{InvariantList}()$
\FOR{each $T \in \mathcal{T}$}
\STATE $\mathcal{X} = \mathcal{X} +$ InvariantGeneration($T$) 
\ENDFOR
\STATE $\mathcal{X}_z$ = Standardize($\mathcal{X}$)  
\STATE $\mathcal{X}_{pc}$ = PCA($\mathcal{X}_z$)  
\STATE $\mathcal{C}$ = Cluster($\mathcal{X}_{pc}$)
\FOR{each $C \in \mathcal{C}$}
\STATE $\mathcal{E}$ = AssignEval($C$) 
\ENDFOR
\FOR{each $T \in \mathcal{T}$}
\STATE $\mathcal{U}$ = ComputeSUS($\mathcal{T}, \mathcal{E}$)
\ENDFOR
\FOR{each $R \in \mathcal{R}$}
\STATE LPRProfile = ConstructProfile($\mathcal{T}, \mathcal{U}$) 
\STATE $L$ = GetMaximalScoringSubsequence(LPRProfile); 
\STATE $\mathcal{L} = \mathcal{L} + L$ 
\ENDFOR
\STATE return $\mathcal{L}$ 
\end{algorithmic} 
\caption{\label{alg:demarcatelinker} 
Linker Demarcation}
\end{algorithm}

We explain each of these steps in greater detail in the rest of the section.

\subsection{Construction and parameterization of LPRs}
Line 1--6 in our algorithm constructs a set of linker probable regions 
(LPRs), $\mathcal{R}$, from the input set of protein structures along with
their domain definitions $S$.  We then represent each LPR i.e. 
$r \in \mathcal{R}$ using a set of overlapping tertrahedrons as described 
below.

Let $\mathcal{S}$ be the set of protein structures along with their domain 
definitions. Each element of $\mathcal{S}$ is an ordered pair of structure 
and its domain definition. Note that since we consider only continuous
multidomain proteins in our analysis, we are in a position to define 
domains using the position of the last amino acid residue in the domain.
We will refer to the position of last amino acid as the endpoint of 
that domain. The set $D$ in ordered pair $(S, D) \in \mathcal {S}$ 
specifies endpoints of each domain in $S$. Thus, for a given structure
$S$ with $e$ domains, $D = \{d_1, d_2, \ldots, d_e\}$. Here $d_j$ is 
the endpoint of domain $j$. Note that the first domain starts at the 
first position and the last domain ends at the last position in the 
protein. Any other domain $j$ with $j > 1$ starts at position $d_{j-1} + 1$
and ends at position $d_j$ in the structure. With this background, we 
are in a position to define LPR. \\ \\
DEFINITION 1: Linker probable region (LPR) between domain $i$ and $j$ of 
protein $s$ is a substructure starting at position $d_i - k + 1$ and 
ending at $d_i + k$ in $s$. It is denoted as $\mathrm{LPR}(s, i, j)$. 
Note that $\mathrm{LPR}(s, i, j)$ contains the end poisition $d_i$ of 
domain $i$ and its length is $2k$. The parameter $k$ is chosen 
based on the average linker length as reported in literature 
~\cite{miyazaki:2002, george:2002a, tanaka:2003}. 
LPR is the basic unit of our analysis.
\\ \\
EXAMPLE 1: Let $s_j$ be the protein structure with two domains. Let 
$D_{s_j} = \{d_1, d_2\}$. $s_j$ has exactly one LPR that starts 
at position $d_1 - k + 1$ and ends at position $d_1 + k$. \\ \\
We construct a set of LPRs, $\mathcal{R}$ by extracting LPRs from 
$\mathcal{S}$, the input set of proteins and their domain definitions 
(line -- 2 in algorithm). Note that all the LPRs in $\mathcal{R}$ are 
of equal length $2k$. The backbone structure of each LPR is 
approximated with its $C_\alpha$ coordinates ~\cite{tendulkar:2004}.
Thus, $R$ is a set of $2k$ amino acid residues along with their 
positions in the structure as given by the $x, y, z$ coordinates. 
Now we will describe a procedure \textsl{DiscretizeLPR} (line -- 3 in 
our algorithm). Each $r \in \mathcal{R}$ is discretized into a sequence of 
$2k - 3$ overlapping tetrapeptides $T = t_{1,2,3,4}, t_{2,3,4,5}, \ldots, 
t_{(2k-3),(2k-2),(2k-1),(2k)}$. Note that the consecutive tetrapeptides 
$t_i$ and $t_{i+1}$ in sequence $T$ share an overlap of three amino acid 
residues. Each tetrapeptide in $T$ is added to $\mathcal{T}$, which is a 
global set of tetrapeptides obtained by discretizing LPRs (line -- 4 in algorithm). 

Each tetrapeptide $t  \in \mathcal{T}$ represents a tetrahedral geometry and 
is described by a fixed suite of $g$ descriptors, which are invariant under 
transformations such as rotation and translation \cite{mumford:1994, weyl:1997, tendulkar:2003, tendulkar:2004}.
These descriptors are referred to as geometric invariants (GIs) in the 
subsequent text. The suite of invariants is carefully chosen after extensive trial and 
error on training data to address the following two issues: 
(a) for superimposable tetrapeptides, the invariants must be 
similar within a tolerance limit $\delta$; and 
(b) for a pair of non-superimposable tetrapeptides $t_1$ and $t_2$, 
there must be at least one geometric invariant such that $f(t_1)$ is 
not similar to $f(t_2)$. Here $f$ is a function that calculates a 
specific GI. We represent each tetrapeptide $t \in \mathcal{T}$ with a suite 
of fifteen GIs (line -- 7). The detailed method for calculating these GIs is 
given in our previous work~\cite{tendulkar:2003, tendulkar:2004}
(line -- 9).
\begin{enumerate}
\item Nine GIs are calculated based on the tetrahedral geometry of $t$ 
and they represent signed volume and perimeter of $t$, length of each edge 
in $t$. Since there are in all six edges so we have six GIs corresponding 
to the length. One more invariant is computed based on the 
sum of distance of each vertex of $t$ from the centroid of all 
the vertices. Let $V_t$ be the set of all vertices in $t$. The $i$-th vertex
$v_i \in V_t$ gives $x, y, z$ coordinate position of $i$-th amino acid 
residue in $t$. The centroid is calculated as follows: 
$$\mu = \frac{1}{n} \sum_{\forall v \in V_t} v$$
and the sum of distance from centroid is calculated as 
$$\sum_{\forall v \in V_t} (v - \mu)$$ 
\item The remaining six invariants for $t$ are calculated by forming three 
triangles using vertices in $V_t$. The three traingles are as follows: 
$v_1, v_2, v_3$, $v_1, v_3, v_4$ and $v_1, v_2, v_4$. We calculate 
area and perimeter for each of these traingles, thus accounting for 
the six remaining invariants. 
\end{enumerate}.

Further we standardize $\mathcal{X}$ to zero mean and unit standard deviation
values. Let $\mathcal{X}_z$ be the set of standardized GIs for $\mathcal{T}$
(Line -- 11). We perform Principal Component Analysis (PCA) to get rid of 
correlations between GIs~\cite{tendulkar:2005}. PCA gives a new set of orthogonal dimensions, 
which are linear combinations of the original dimensions (GIs in this case). 
We selected first $m$ significant principal components (PCs) to represent 
the tetrapeptides. Let $\mathcal{X}_{pc}$ be the set of tetrapeptides represented
using $m$ PCs (Line -- 12). 

\subsection{Structural profiling of LPRs}
Since the linkers are unstructured regions, we believe that the corresponding 
tetrapeptides share structural similarity with fewer other tetrapeptides. On the 
other hand, the tetrapeptides from non-linker region are expected to share 
structural similarity with a large number of other tetrapeptides. Our objective 
is to determine the groups of structurally similar tetrapeptides through 
clustering process and utilize this knowledge towards the demarcation of 
actual linkers. We perform clustering of a set of tetrapeptides $\mathcal{X}_{pc}$,
which are represented using $m$ PCs as explained earlier (Line -- 13). 
The clustering is carried out via Matlab implementation of hierarchical 
agglomerative clustering (HAC) \cite{kumar:2005}. We use euclidean distance 
as a measure of similarity and ward linkage \cite{kumar:2005} for merging the 
nodes in the clustering tree. The optimal cut in the resulting dendrogram is 
determined by using inconsistency parameter, which leads to the discovery of 
a set of clusters $\mathcal{C}$. The inconsistency parameter compares each 
link in the cluster hierarchy with the adjacent links to determine natural 
cluster division in the dataset \cite{kumar:2005}. The clustering process 
assigns each tetrapeptide to exactly one cluster.

Once the clustering process is over, we obtain the distribution of cluster 
sizes, which is used to assign e-value to each cluster based on its size (Lines 14--16). 
The e-value for a cluster $C \in \mathcal{C}$ with size $|C|$ is calculated
as $\alpha/|\mathcal{C}|$, where $\alpha$ be the number of clusters in $\mathcal{C}$ 
with size greater than $|C|$ and $|\mathcal{C}|$ is the total number of clusters.
Note that the large clusters are expected to contain tetrapeptides corresponding 
to the non-linker regions, while the smaller clusters are more likely to 
contain tetrapeptides corresponding to the linker region.
The e-values are normalized to zero mean and unit standard deviation to 
obtain structural uniqueness score (SUS) of the cluster. The large clusters 
have lower SUS, while the smaller clusters have higher SUS. The smallest 
SUS is assigned to the largest clusters, while the largest SUS is assigned 
to the singleton clusters. Thus, the SUS indicates the structural uniqueness 
of the cluster and its propensity to be a part of the actual linker. Each 
tetrapeptide in the cluster is assigned the SUS of that cluster (Lines 17--19). The 
structural profile of an LPR is represented using the SUS of its 
constituent tetrapeptides (Line 21).

\subsection{Protein domain linker demarcation}
Given the structural profile of an LPR, we are interested in finding the longest 
continuous stretch of tetrapeptides with the highest cumulative SUS. Note that 
such tetrapeptides; being highly unstructured; appear as outliers in the 
clustering process. Hence, the stretch is demarcated as a linker between the 
domains. We are required to enumerate all possible stretches in order to find 
the one with the greatest cumulative SUS. The problem is tackled by using 
linear time dynamic programming algorithm proposed by Ruzzo and Tompa 
~\cite{ruzzo:1999} (Line -- 22). The algorithm takes a sequence of real 
numbers as input and generates non-overlapping, contiguous subsequences 
having greatest total score. Here, the algorithm takes the structural 
profile of an LPR as input, which is a sequence of nine SUS scores of 
the constituent tetrapeptides $(u_1, u_2, \ldots , u_{2k-3})$, 
where $u_i \in R$. Let $Q$ be the set of all possible subsequences
of tetrapeptides. The cumulative SUS for each subsequence is obtained by 
simply summing the SUS of the constituent tetrapeptides. Let $\mathrm{\texttt{CumSUS}} (q)$ be 
the function that gives cumulative score for the subsequence $q \in Q$. 
The \textsl{GetMaximalScoringSubsequence} procedure finds the subsequence
with the greatest cumulative SUS (Line -- 22).  We declare such a subsequence as a 
domain linker. In case of a tie, a subsequence with the closest proximity to 
the domain boundary is declared as a domain linker. Thus,
$$
L = \mathrm{argmax}_{q \in Q} \mathrm{\texttt{CumSUS}}(q) 
$$

\subsection{Evaluation of proposed method}
It is of interest to evaluate the accuracy of the proposed method. In the absence
of linker database and due to a lot of subjectivity in linker detection by 
visual examination, we decided to extract experimentally reported linkers 
from the literature. We first selected research papers based on PDB reference 
record of each protein in our input dataset. We then manually read the 
literature for extracting information about experimentally detected linkers. 
We succeded in extracting linker information about 51 proteins out of 725 
proteins in the input set (Supplementary Table 1). These linkers form an evaluation set 
for the benchmark studies.  

After demarcating the linkers using the proposed method, we compare them with the 
literature reported linkers. We compute the accuracy of demarcation residue wise
as follows: If the residue marked as a part of the linker also happens to be 
the part of the literature reported linker, we count it as a true positive 
match, else it is counted as a false positive match. If the residue that is part
of the literature reported linker, but is not present in the linker marked by
the proposed method, it is counted as a false negative match. Let $TP$ denotes 
the number of correctly demarcated linker residues, $FP$ denotes the number 
of incorrectly demarcated linker residues, which are actually non-linker 
residues and $FN$ denotes the number of actual linker residues, which were not 
included in demarcated linker region. Based on $TP, TN$ and $FN$, we compute 
precision and recall of the proposed method as follows:
\begin{eqnarray*}
\mathrm{Recall} = \frac{TP}{TP+FN} \\
\mathrm{Precision} = \frac{TP}{TP+FP} 
\end{eqnarray*}
We also compute $F1$ measure, which is harmonic mean of precision and recall,
for the proposed method.

\section{Results}
\subsection{Dataset preparation and structural profiling of LPRs}
We have selected 610 continuous multi-domain proteins from the 
ASTRAL 40 \cite{Brenner:2000} dataset version 1.69. Out of 610 
selected proteins, we have 505 two domain, 95 three domain and 10 
four domain proteins. Based on the SCOP \cite{Murzin:1995} domain 
definition, we selected a stretch of 6 amino acids on either side of the 
domain boundary point to extract LPR of length 12 for each domain connection.
Thus, we obtain 725 LPRs from the input protein domains. Each LPR is 
represented by nine overlapping tetrapeptides with an overlap of three 
residues between the consecutive tetrapeptides. Each tetrapeptide is 
represented with fifteen geometric invariants (GIs) as described earlier. 
Thus, we obtain 6525 tetrapeptides represented in 15 dimensional space 
spanned by GIs. This dataset is subjected to PCA, which reveals that 
the first 8 PCs cover 99\% variance in the data. The tetrapeptides were 
then transformed into a reduced dimensional space spanned by first 
8 PCs. The transformed dataset of tetrapeptides is subjected to hierarchical 
clustering algorithm (Matlab implementation). The resulting dendrogram was 
cut based on inconsistency parameter to obtain 2188 clusters. The 
distribution of clusters in terms of their size is shown in Table 1. Note 
that we obtain a large number of smaller clusters, approximately 50\%, with 
size less than three members. The largest cluster contains 14 tetrapeptides. 

\begin{table}[h]
\begin{center}
\begin{tabular}{cc}
\hline 
\textbf{Cluster Size} & \textbf{Number of Clusters} \\ \hline
1 & 207 \\ 
2 & 899 \\ 
3 & 520 \\ 
4 & 269 \\ 
5 & 131 \\ 
6 & 58 \\ 
7 & 45 \\ 
8 & 23 \\ 
9 & 13 \\ 
10 & 9\\ 
11 & 3\\ 
12 & 4\\ 
13 & 4\\ 
14 & 3\\ \hline
\end{tabular}
\caption{\footnotesize Distribution of clusters of tetrapeptides according to their size.}
\end{center}
\end{table}
The larger clusters are assigned smaller e-values, while the smaller 
clusters are assigned larger e-values. We then constructed the structural 
profile of LPRs using the SUS of the corresponding tetrapeptides. The structural 
profiles, each of length nine, were subjected to a maximally scoring 
subsequence finding algorithm to demarcate the actual linkers. We were able to 
demarcate 692 domain linkers from 725 input LPRs.	In the remaining 33 cases, 
we observed that these LPRs contain tetrapeptides with lower SUS.	The 
distribution of linker lengths is shown in Table 2.	We found that the 
average length of the linker detected by the proposed method is 
5.3 residues. 
\begin{table}[h]
\begin{center}
\begin{tabular}{|c || c c c c c c c c|}
\hline 
\textbf{Linker Length} & 4 & 5 & 6 & 7 & 8 & 9 & 10 & 11 \\ \hline
\textbf{Linker Count} & 286 & 178 & 93 & 67 & 31 & 19 & 11 & 7 \\ \hline
\end{tabular}
\caption{\footnotesize Distribution of linkers according to their lengths. The linkers
are obtained via the proposed method}
\end{center}
\end{table}

\subsection{Comparison with other methods}
We were interested in comparing the proposed method against the state of the 
art methods to assess its performance. We used 51 literature reported 
linkers for the comparative analysis. The same set was used for evaluating the 
proposed method. We have selected the following methods in the comparison study: 
Ebina et al.~\cite{ebina:2009}, GM\cite{galzitskaya:2003} and 
CHOP~\cite{liu:2004}. Note that the direct comparison is inappropriate 
since most of these methods predict putative domain linkers from the sequence 
characteristics, while our method demarcates domain linkers by analyzing 
structural characteristics of LPRs. Moreover most of these methods predict 
multiple putative linkers with certain flexibility on start and end positions. 
From these predictions, we selected the most appropriate linker based on the 
known domain definition and used it for the comparison. A representative examples of 
the linkers identified by different methods on the input set are reported in 
Table 3. The complete list can be obtained from Supplementary Table 1. 
These predictions are matched with the actual linkers and the accuracy is calculated 
in terms of F1 score, which is the harmonic mean of precision and recall. The comparative 
performance of the proposed method is given in Table 4. The proposed method 
achieves overall recall of 0.66 and precision of 0.83 on the benchmark dataset. 
It significantly outperforms state of the art methods in terms of the number of linkers 
identified as well as the accuracy of the predictions. 

\begin{table}[h]
\begin{tabular}{p{0.5in}p{0.4in}p{0.5in}p{0.7in}p{0.5in}p{0.5in}p{0.5in}}
\hline
\textbf{PDB ID}&\textbf{Lit. Ref.}&\textbf{Actual Linker}&\textbf{Proposed Method}&\textbf{Ebina et. al.~\cite{ebina:2009}}&\textbf{GM~\cite{galzitskaya:2003}}&\textbf{CHOP~\cite{liu:2004}}\\\hline
1h03&\cite{williams:2003}&65-68&64-70&47-73&50-70&66-67\\
1eqf&\cite{Jacobson:2000}&1495-1502&1492-1503&1521-1540&1504-1524&1496-1515\\
1fcd&\cite{chen:1994}&76-84&77-80&75-83&97-117&78-79\\
1vi7&\cite{park:2004}&134-139&133-138&132-138&156-176&137-138\\
1fp5&\cite{Wurzburg:2000}&436-440&438-442&420-461&436-456&446-447\\
1fx7&\cite{Feese:2001}&141-150&143-146&139-147&137-157&139-151\\
1f1b&\cite{Jin:2000}&95-101&95-102 &93-102&88-108&100-101\\
1f14&\cite{Barycki:2000}&201-206&201-204&193-206&249-269&218-222\\
1jt6&\cite{Schumacher:2001}&72-74&70-73&73-76&123-143&51-52\\
1m3y&\cite{Nandhagopal:2002}&213-224&220-226&213-226&244-264&223-224\\
1eem&\cite{Board:2000}&98-107&100-104&97-112&82-102&125-126\\
1g3n&\cite{Jeffrey:2000}&142-150&149-152&133-136&193-213&150-151\\
1l3l&\cite{Zhang:2002}&163-174&168-172&166-170&146-166&171-172\\
1gnw&\cite{Reinemer:1996}&78-92&84-87&106-113&107-127&88-92\\
1dkz&\cite{Zhu:1996}&503-508&501-512&535-540&459-479&541-542\\
1e79&\cite{Gibbons:2000}&95-105&96-101&66-80&61-81&117-118\\
1gpj&\cite{Moser:2001}&142-148&141-144&136-142&153-173&142-146\\
1hf2&\cite{Cordell:2001}&96-102&95-99&82-87&132-152&100-102\\
1hlv&\cite{Tanaka:2001}&65-74&64-67&56-57&70-90&66-72\\
1hv8&\cite{Story:2001}&208-214&207-210&249-253&300-320&208-211\\
1hyr&\cite{Li:2001}&175-184&178-181&175-194&185-205&183-184\\
1jb9&\cite{Aliverti:2001}&156-164&158-161&159-164&155-175&158-159\\
1jbw&\cite{Sun:2001}&295-300&291-300&278-288&345-365&298-299\\
1k0d&\cite{Bousset:2001}&197-205&195-198&175-186&195-215&210-211\\
1k0m&\cite{Harrop:2001}&89-100&92-97&90-95&96-116&97-98\\\hline
\end{tabular}
\caption{\footnotesize The table contains a representative examples of 
linkers extracted by the proposed method. We have also shown actual linker
as extracted from the literature as well as the linkers predicted by 
state of the art methods. The column \textsl{Lit. Ref.} provides
the literature reference for the actual linker.}
\end{table}

\begin{table}[h]
\begin{tabular}{l|c|c|c|c}
\hline
{\bf Method} & {\bf Precision} & {\bf Recall} & {\bf F1 Score} & {\bf No. of Predictions} \\ \hline
Proposed Method & {\bf 0.83} & {\bf 0.66} & {\bf 0.74} & {\bf 51} \\
Ebina et. al.~\cite{ebina:2009} & 0.48 & 0.57 & 0.52 & 29 \\
CHOP~\cite{liu:2004} & 0.35 & 0.39 & 0.37 & 36 \\
GM~\cite{galzitskaya:2003} & 0.19 & 0.50 & 0.27 & 22 \\ \hline
\end{tabular}
\caption{\footnotesize Performance of various methods on the benchmark dataset of 51 linkers}
\end{table}

We further compared our method against DomCut, which predicts the domain cut 
point based on domain linker index.	Note that DomCut does not predict the 
start and the end position of linker. The DomCut prediction is taken as a 
correct prediction if the predicted domain cut point falls within the actual 
linker. Out of 51 linkers, we found that DomCut predicts correctly in 13 cases 
and does not predict the domain cut point in 14 cases. In the remaining 
cases, the predicted cut point does not fall inside the actual linker.

We computed the agreement between our method against the linker database of 
George and Heringa~\cite{george:2002, george:2002a}, which gives linker predictions for 79 proteins in 
our input dataset. Note that linker prediction is available for few proteins 
from evaluation set used earlier and hence it is not used for the comparison 
between the two methods. We found that the predictions partially agree in 55 
cases and completely disagree in 24 cases. Reasonable agreement ($>75\%$) was 
obtained in 5 cases, while medium agreement ($< 75\%$ and $> 40\%$) was obtained 
in 29 cases and weak agreement was observed in the remaining cases.

\subsection{Representative linkers}
The examples of demarcated linkers by the proposed method are shown in 
Figures 2A–2J. Since the method use LPR for demarcation, the entire 3-D 
structure of corresponding domains is not shown. Instead, a stretch of 16 
amino acids on either side of the domain boundary point is shown to 
maintain clarity of representation. Here we describe a few representative 
linkers. \\ \\
\textsl{Example of literature reported linkers, also demarcated by 
our method}
\begin{enumerate}
\item \textsl{Streptococcus pneumonia SP14.3 (PDB Code: 1IB8)} \\
Streptococcis pneumonia is a deadly human pathogen causing high mortality 
and morbidity rates ~\cite{yu:2001}. SP14.3 is a key protein responsible for 
growth of the pathogene. The three-dimensional structure of SP14.3 contains a 
very short linker of size 3 between residues 88-90 with a moderate flexibility. 
Our method predicts the linker exactly at the same place as reported by 
Yu et. al. ~\cite{yu:2001}. The linker plays role in relative orientation of 
domains and maintaining rotational cooorelation of domains. 
\item \textsl{Yeast Sec18p (PDB Code: 1CR5)} \\
Yeast Sec18p is a hexameric ATPase with a central role in vesicle trafficking 
\cite{babor:1999}. 
The reported linker is located between residues 104-113 and is flexible in its 
structure. The SNAP binding site is located opposite to the linker. It connects 
two beta rich sub domains and is likely to facilitate different sub-domain 
orientation. Our method detected the linker between residues 104-111, 
which is enclosed within the literature reported linker (Figure 2A). 
\item \textsl{TnsA (PDB Code: 1F1Z)} \\
TnsA carries out DNA breakage at 5’ end of transposon. It contains a six sized loop 
linker between residues 165-170 that connects two domains of homodimeric 
endonuclease enzymes~\cite{hickman:2000}. Our method predicted eight sized 
linker between residues 164-171 (Figure 2B). The linker is likely to play a role 
in cooperative domain binding.
\end{enumerate}
\textsl{Examples of novel linkers that are not reported in literature}\\
Figure 2G shows the novel linker in Ornithine transcarbamoylase (1DUV) which 
is not reported in the literature. Fascin (1DFC) has three domains, and the 
two linkers delimiting these domains were demarcated accurately (Figures 2H and 2I). 
Spectrin beta chain (1S35) is an all alpha protein with an alpha-helical 
linker between two domains (Figure 2J). A helical conformation of the linker 
region is compatible with a variety of different twist angles between Spectrin 
repeats, leading to many different conformations. 
\begin{figure}[th]
\centering
\includegraphics[scale=1.5]{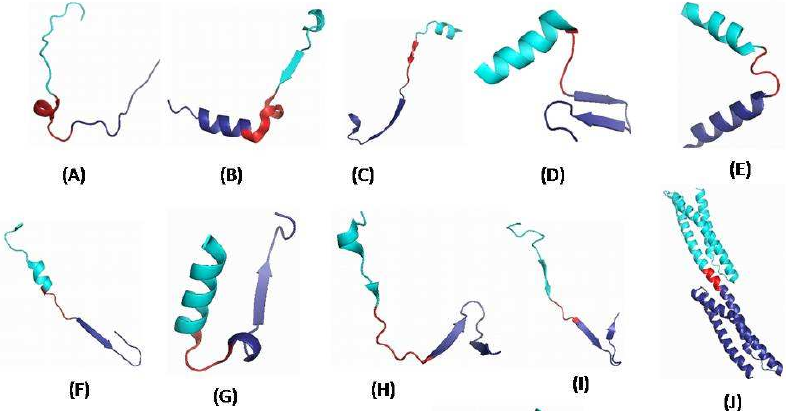}
\caption{\label{fig:linkers}\footnotesize Different categories of substructures predicted as 
linkers. The segment of the N- terminal domain is shown in cyan whereas the 
segment of the C-terminal domain is shown in blue. The segment of the 
structure demarcated as a linker is shown in red. (A) 1CR5 [N--terminal 
domain of Sec18p], helical linker [LEU 104 -– GLN 111]; (B) 1F1Z [TnsA endonuclease], 
linker [THR 164 -– VAL 171]; (C) 1VI7 [Hypothetical protein YigZ], beta 
linker [THR 133 –- PRO 138]; (D) 1P2F [Response regulator DrrB from 
Thermotoga maritime], loop linker [GLU 118 -– GLY 121]; (E) 1R89 
[tRNA nucleotidyltransferase], loop linker [GLY 139 –- GLY 143]; 
(F) 1DT9 [Eukaryotic peptide chain release factor subunit 1], novel 
linker [LEU 140 -– SER 144]; (G) 1DUV [Ornithine transcarbamoylase], 
novel linker [LEU 148 –- ALA 152]; (H) 1DFC [Fascin], loop linker between 
domain 1 and domain 2 [HIS 1135 -– GLN 1141]; (I) 1DFC [Fascin], loop 
linker between domain 2 and domain 3 [SER 1259 -– GLN 1262]; 
(J) 1S35 [Spectrin beta chain], helical linker [THR 1163 – PHE 1170]. 
The figures are prepared using PyMOL (http://www.pymol.org).}
\end{figure}

\section{Discussion}
We proposed a novel objective method for accurate demarcation of linkers. The 
method takes three dimensional protein structure and domain definition as an 
input and provides accurate linker demarcation. This is the first instance 
where structural aspects are rigorously analyzed in the linker demarcation 
task. The earlier methods have used biochemical and sequence properties for 
the same task. Since the proposed method provides structural perspective in 
the demarcation, it can be used in tandem with the other methods reported 
in literature.

As stated earlier, accurate domain linker demarcation is a key to understand 
their biochemical properties. Given a three dimensional structure and its 
domain definition, the linkers can be demarcated either through direct 
visualization or through objective automated methods reported in the 
literature \cite{george:2002, miyazaki:2002, suyama:2003, bae:2005, dumontier:2005, 
ebina:2009}. Visualization methods are often subjective, while 
automated methods demarcate linkers only approximately. The proposed 
method demarcates the linker more precisely than the other methods as 
demonstrated on the benchmark dataset. Our method outperforms other 
methods with F1 score of $0.745$ (precision 0.83 and recall 0.66) 
on the benchmark dataset.

The method is also the first of its kind in exploiting biological knowledge 
about structural uniqueness of linkers. Since the linkers possess flexible 
structure, their constituent fragments are unique and appear as outliers during 
clustering process \cite{wriggers:2005}. 
Since the outliers are assigned the maximum SUS, the stretch of fragments with maximum 
cumulative SUS corresponds to the actual linker. The discovery of such stretch 
was performed using an efficient linear time dynamic programming algorithm.

The proposed method has following configurable parameters, which can be adjusted 
to achieve desired results: (i) $k$ which affects the length of LPRs and 
(ii) the length of the peptide fragment. The current study uses LPRs of length 12, 
tetrahedron as a choice for local structure and clusters of tetrahedrons in LPRs. 
The length of LPR was decided based on the prior reports of average linker lengths 
\cite{george:2002, miyazaki:2002, tanaka:2003}.
Flexibility can be added to LPR selection with the help of other methods 
reported in the literature. For instance, we can use Taylor’s method 
\cite{taylor:1999} as applied by \cite{george:2002} to come up with more 
appropriate LPRs. The LPRs thus obtained can be processed further with the help 
of the proposed method to demarcate accurate linkers. In the present study, we 
extract tetra-peptides from LPRs and perform the clustering. The clusters are 
then assigned SUS based on cluster size distribution. Our method is able 
to find linkers with irregular or unique structure. It is also able 
to detect linkers containing alpha-helices and beta-strands with structural 
purturbations.
Due to these purturbations, these tetrapeptides 
are part of smaller clusters, which are often assigned higher SUS and hence 
our method is able to detect linkers containing such structures.
However, our method is unable to detect linkers made up of regular alpha-helices
and beta-strands, since the regular structures tend to form larger clusters 
and usually have smaller SUS compared to the irregular structures. 

Finally, we plan to construct a database of linkers demarcated via the proposed 
method. The database will help to further our understanding of biochemical 
properties of linkers and help to design better linkers while engineering 
multi-domain proteins. We are planning to use insights about linkers from 
this work to develop a sequence based linker prediction method. This can be 
useful in predicting protein domains by virtue of linker prediction.

\section{Acknowledgement}
We are grateful to Prof. Pramod Wangikar - IIT Bombay and Mr. Nivruti Hinge for 
their useful discussions and timely help. We thank Michael Tress for his useful 
comments on the manuscript. This work is supported by Innovative Young Biotechnologist
Award (IYBA) grant from Department of Biotechnology of Government of India to AVT.

\bibliographystyle{abbrv}
\bibliography{manuscript}  
\end{document}